\documentclass{article}
\usepackage{jamaica04}
\usepackage{epsfig}
\frompage{000} \topage{000}                                              
\def\rightmark{Collective flow with PHOBOS}
\def\leftmark{S. Manly et al.}

\title{Collective flow with PHOBOS}
\authors{
{S.Manly$^{8}$ for the PHOBOS Collaboration\\
B.B.Back$^1$,
M.D.Baker$^2$, M.Ballintijn$^4$, D.S.Barton$^2$, B.Becker$^2$,
R.R.Betts$^6$, A.A.Bickley$^7$, R.Bindel$^7$, A.Budzanowski$^3$,
W.Busza$^4$, A.Carroll$^2$, M.P.Decowski$^4$, E.Garc\'{\i}a$^6$,
T.Gburek$^3$, N.George$^{1,2}$, K.Gulbrandsen$^4$, S.Gushue$^2$,
C.Halliwell$^6$, J.Hamblen$^8$, A.S.Harrington$^8$,
G.A.Heintzelman$^2$, C.Henderson$^4$, D.J.Hofman$^6$,
R.S.Hollis$^6$, R.Ho\l y\'{n}ski$^3$, B.Holzman$^2$,
A.Iordanova$^6$, E.Johnson$^8$, J.L.Kane$^4$, J.Katzy$^{4,6}$,
N.Khan$^8$, W.Kucewicz$^6$, P.Kulinich$^4$, C.M.Kuo$^5$,
J.W.Lee$^4$, W.T.Lin$^5$, S.Manly$^8$, D.McLeod$^6$,
A.C.Mignerey$^7$, R.Nouicer$^{2,6}$, A.Olszewski$^3$, R.Pak$^2$,
I.C.Park$^8$, H.Pernegger$^4$, C.Reed$^4$, L.P.Remsberg$^2$,
M.Reuter$^6$, C.Roland$^4$, G.Roland$^4$, L.Rosenberg$^4$,
J.Sagerer$^6$, P.Sarin$^4$, P.Sawicki$^3$, I.Sedykh$^2$,
W.Skulski$^8$, C.E.Smith$^6$, P.Steinberg$^2$, G.S.F.Stephans$^4$,
A.Sukhanov$^2$, J.-L.Tang$^5$, M.B.Tonjes$^7$, A.Trzupek$^3$,
C.Vale$^4$, G.J.van~Nieuwenhuizen$^4$, R.Verdier$^4$,
G.I.Veres$^4$, F.L.H.Wolfs$^8$, B.Wosiek$^3$, K.Wo\'{z}niak$^3$,
A.H.Wuosmaa$^1$, B.Wys\l ouch$^4$, J.Zhang$^4$
}\\[2.812mm]
{\normalsize \hspace*{-8pt}$^1$~Argonne National Laboratory,
Argonne,
IL 60439-4843, USA\\[0.2ex]
\hspace*{-8pt}$^2$~Brookhaven National Laboratory, Upton, NY
11973-5000, USA\\[0.2ex]
\hspace*{-8pt}$^3$~Institute of Nuclear Physics PAN, Krak\'{o}w, Poland\\[0.2ex]
\hspace*{-8pt}$^4$~Massachusetts Institute of Technology,
Cambridge, MA 02139-4307, USA\\[0.2ex]
\hspace*{-8pt}$^5$~National Central University, Chung-Li, Taiwan\\[0.2ex]
\hspace*{-8pt}$^6$~University of Illinois at Chicago, Chicago,
IL 60607-7059, USA\\[0.2ex]
\hspace*{-8pt}$^7$~University of Maryland, College Park, MD 20742, USA\\[0.2ex]
\hspace*{-8pt}$^8$~University of Rochester, Rochester, NY 14627,
USA}}

\abstract{This paper reviews recent results on directed and
elliptic flow from the PHOBOS experiment using data taken during
Au+Au runs at RHIC. The systematic dependence of flow on
pseudorapidity, energy, transverse momentum and centrality is
discussed. }


\begin{document}

\maketitle
\setcounter{page}{1}

\section{Introduction}\label{intro}

Collective flow has proven to be one of the more fruitful probes
of the dynamics of heavy ion collisions at RHIC. The elliptic flow
signal (v$_{2}$) at mid-rapidity is large and consistent with
expectations from hydrodynamic models at low p$_{T}$\cite{starpt}.
It has been interpreted as evidence for the production of a highly
thermalized state, and perhaps for partonic matter
\cite{gyulassy}. At high p$_{T}$, the observed suppression of
elliptic flow \cite{starjq1,starjq2} is consistent with
calculations incorporating jet quenching \cite{JQ} and quark
coalescence \cite{QC}. Interestingly, the fall of v$_{2}$ with
pseudorapidity ($|\eta|$) has been less amenable to understanding
\cite{hirano}.

This paper provides an update on the status of flow studies by the
PHOBOS collaboration using data from Au+Au collisions at RHIC. The
data were recorded in the years 2000 ($\sqrt{s_{_{NN}}} =$130 GeV)
and 2001 ($\sqrt{s_{_{NN}}} =$19.6 and 200 GeV). Given the wide
range of pseudorapidity coverage and energies present in the data,
it is
interesting to examine the extent to which the shape of the flow
distributions change with energy in the frame of reference of one
of the incoming nuclei. In this paper, the term ``limiting
fragmentation" will be used to describe the extent to which energy
independence holds in this frame of reference.  This usage may
extend well beyond the region of the collision normally thought of
as the fragmentation region.

\section{PHOBOS flow measurement techniques}\label{techno}

The PHOBOS experiment employs silicon pad detectors to perform
tracking, vertex detection and multiplicity measurements. Details
of the setup and the layout of the silicon sensors can be found in
reference \cite{phobos_det}.

PHOBOS flow analyses to date have made use of the subevent
technique described in reference \cite{pandv}. Results from three
independent methods are shown in this paper. The two ``hit-based"
analyses make use of position information (only) for tracks
traversing the octagon and ring subdetectors which cover $-5.4 <
\eta < 5.4$ \cite{phflow}.  The ``track-based" analysis includes,
in addition, reconstructed track (momentum) information from the
spectrometer subdetector which covers $0 < \eta < 1.5$
\cite{qm02}.

The two hit-based analyses differ primarily in the selection of
the event collision point.  In the first method, known as the
offset vertex method, interactions are chosen such that they are
centered in an azimuthally symmetric part of the detector. The
second method, called here the full acceptance method, makes use
of interactions centered near the nominal interaction point in the
azimuthally asymmetric parts of the detector. These azimuthal
asymmetries are symmetrized in software in the early stages of the
analysis. The earlier PHOBOS flow results were obtained with the
former technique \cite{phflow}. However, the latter method is more
compatible with the event trigger used for the bulk of our data
and is the only available technique for data taken at
$\sqrt{s_{_{NN}}}=19.6$ GeV. Good agreement has been shown between
PHOBOS elliptic flow measurements using hit-based and track-based
methods, even though the susceptibility of the two techniques to
background effects is quite different~\cite{qm02}.

\subsection{Results}\label{results}

\begin{figure}[ht]
\begin{center}
\includegraphics[height=4.7cm]{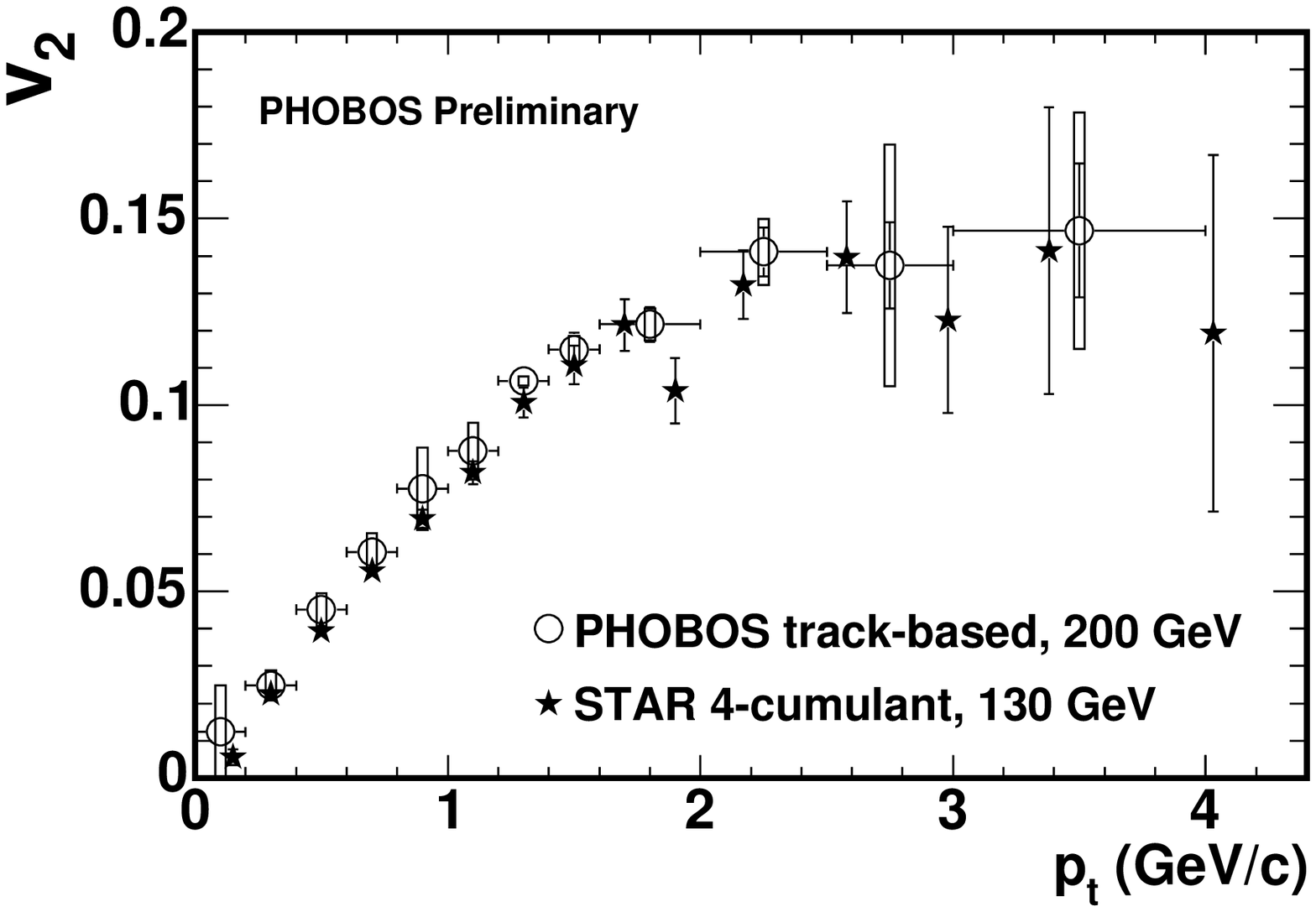}
\includegraphics[width=5.6cm]{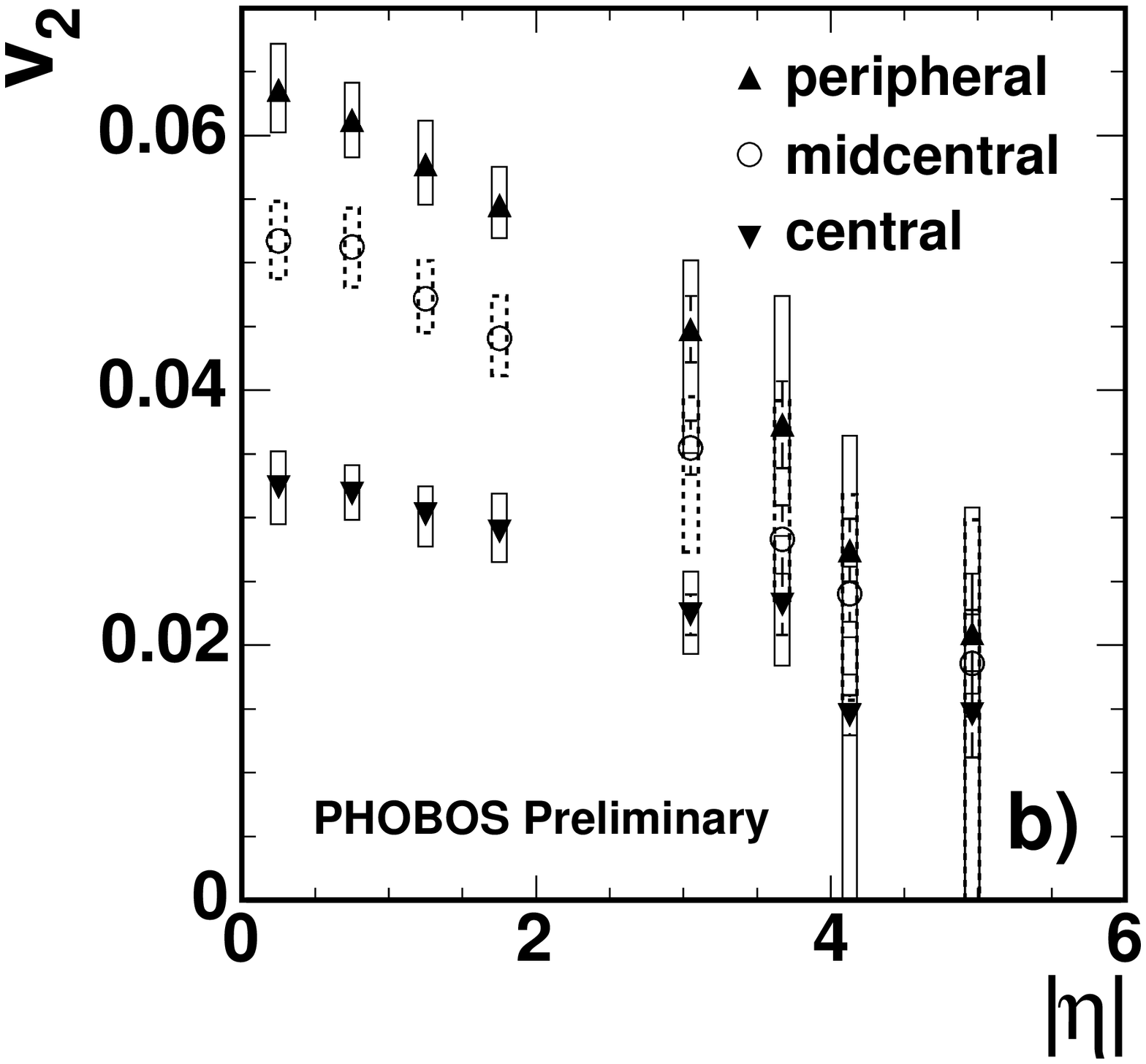}
\end{center}
\caption{\label{v2figs} a) Comparison of PHOBOS (circles) $v_{2}$
for 0-55\% centrality Au+Au collisions to STAR Au+Au collisions at
5-53\% centrality. STAR results using the 4-particle cumulant
(stars) method are shown \cite{starcumulant}. PHOBOS error bars
are statistical, with error boxes representing systematic
uncertainty at a 90\% confidence level(CL). b) $v_{2}$ vs.
$|\eta|$ at different centralities: peripheral (25-50\%),
midcentral(15-25\%), and central(3-15\%) for 200 GeV Au+Au
collisions. Statistical error bars are smaller than the data
points, and boxes show systematic errors at 90\% CL.}
\end{figure}

The charged hadron $v_{2}$ as a function of $p_{t}$, measured by
PHOBOS for 200 GeV Au+Au collisions is shown in
Figure~\ref{v2figs}a). Also shown are STAR data for 130 GeV Au+Au
collisions analyzed with the 4-particle cumulant technique.
Previous measurements demonstrate that elliptic flow changes
little from 130 to 200 GeV at $p_{T}> 1$ GeV/c \cite{v2match}. The
PHOBOS measurement is very similar to STAR's 4-particle cumulant
results, which has been shown to be insensitive to non-flow
effects \cite{starcumulant}. This confirms our expectation that
the PHOBOS track-based flow results are largely unaffected by
non-flow correlations.  It was also determined that the hit-based
and track-based results agree extremely well, implying that the
hit-based results are also free from large non-flow effects.

Figure~\ref{v2figs}b) shows the pseudorapidity dependence of the
elliptic flow for 200 GeV Au+Au collisions for three broad
centrality bins.  In this plot, the hit-based and track-based
results are combined. The peripheral data do not appear to be
flat, even at midrapidity. Within the uncertainties, the shape of
$v_{2}(|\eta|)$ is not strongly centrality dependent, appearing to
differ by only a scale factor.

\begin{figure}[htb]
\centerline{ \epsfig{file=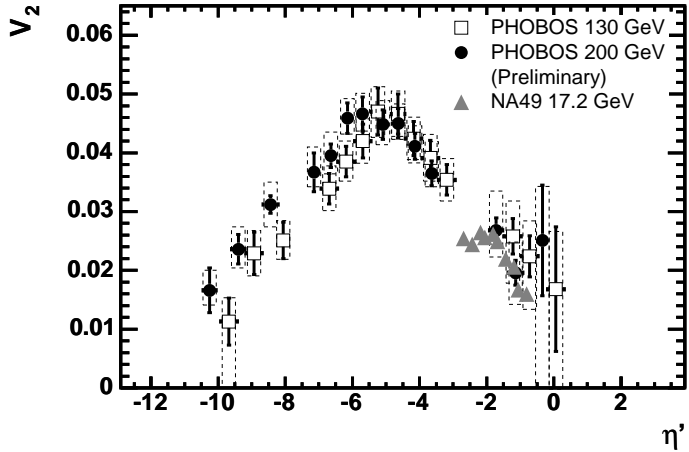,width=10cm} }
\caption{Elliptic flow at three energies as a function of $\eta
'=\eta-y_{beam}$. The 200 \cite{qm02} and 130 \cite{phflow} GeV
data are previously published Au+Au results from PHOBOS. The
17.2-GeV points represent Pb-Pb data (for pions) from NA49
\cite{na49}.} \label{v2etaprime}
\end{figure}

The elliptic flow at three energies, plotted in the frame of
reference of one of the beam nuclei as a function of $\eta
'=\eta-y_{beam}$, is shown in Figure~\ref{v2etaprime}.
Note the remarkable agreement
in the curves all the way from $\eta '=0$ to the mid-rapidity
region for each energy. The results are qualitatively unchanged if
one factors out the $\sim$10\% peaking in v$_{2}$ near
mid-rapidity expected due to the fact that the results are
determined in $\eta$ rather than in y \cite{kolb}.

\begin{figure}[ht]
\begin{center}
\includegraphics[width=5.7cm]{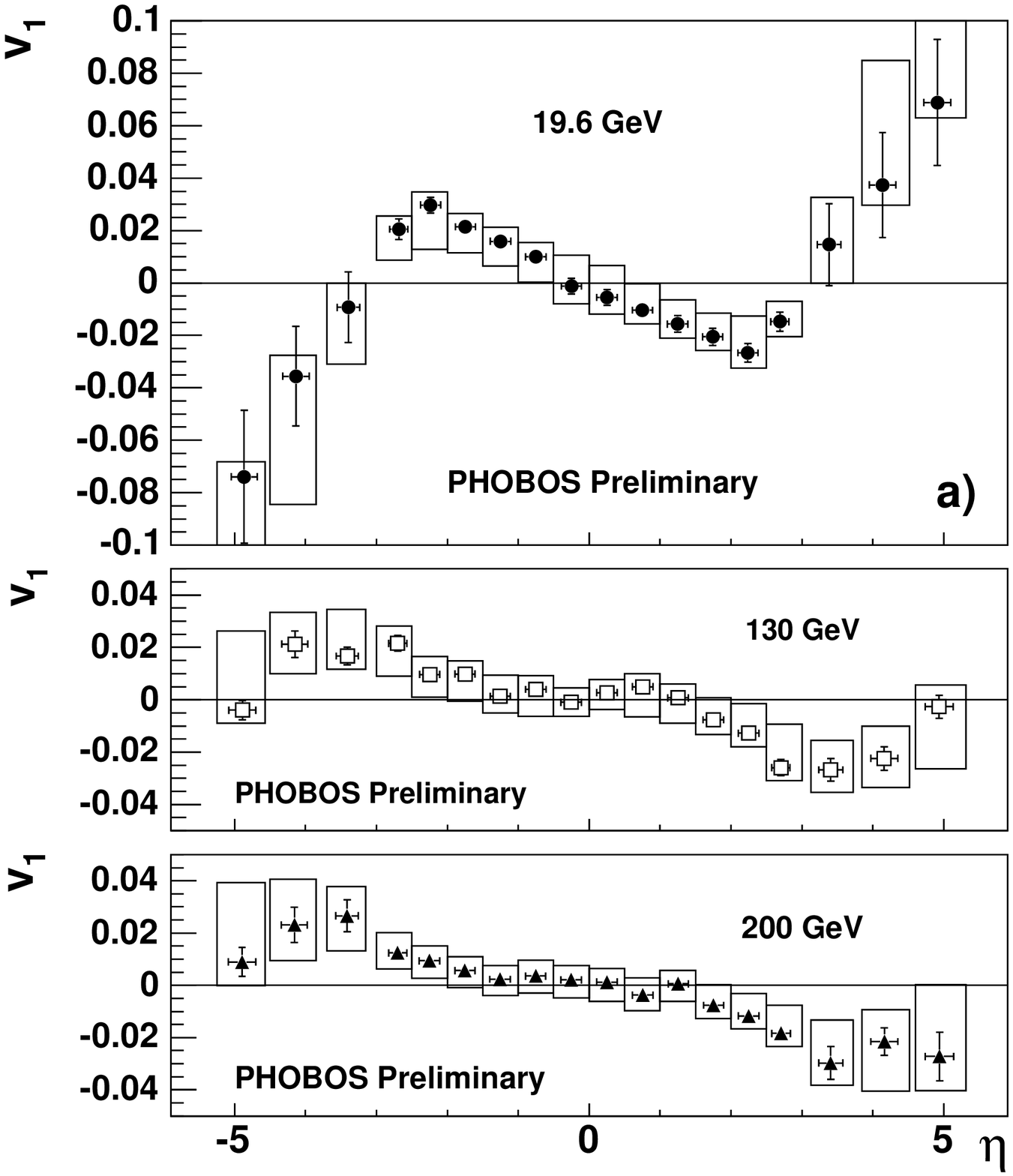}
\includegraphics[width=5.7cm]{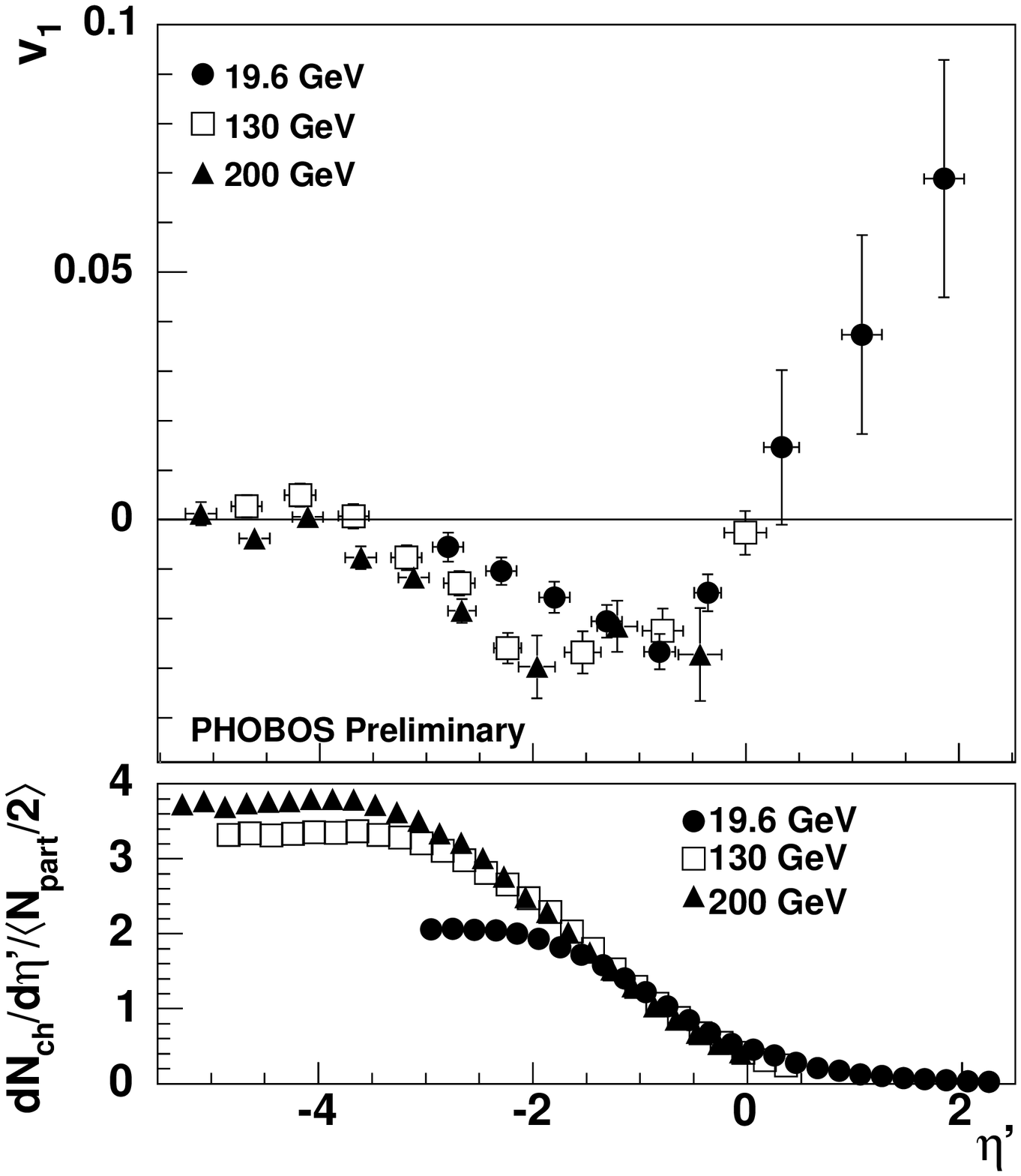}
\end{center}
\caption{(a) The directed flow measured as a function of $\eta$
for $\sqrt{s_{NN}}=$ 19.6 (top), 130(middle), and 200 GeV(bottom)
Au+Au collisions. Error bars are statistical, the height of boxes
represents the systematic error, and the width of the boxes marks
the size of the $\eta$ bin. (b) Directed flow, as shown in (a),
but translated to $\eta '=\eta-y_{beam}$. Systematic error boxes
are omitted for clarity. (c) Scaled pseudorapidity distributions
for 0-6\% central 19.6, 130, and 200 GeV Au+Au collisions as a
function of $\eta '$ \cite{limfrag}. Systematic errors not shown.}
\label{v1stack}
\end{figure}

The PHOBOS results on directed flow ($v_{1}$) are measured with
the full acceptance hit-based method \cite{qm04}. The flow is
calculated for hits measured in the octagon with an event plane
determined from widely separated subevents (constructed to be
symmetric in $\eta$). Subevents in the octagon are used for flow
measured in the rings. Figure \ref{v1stack}a) shows $v_{1}$ vs.
$\eta$ for charged hadrons measured in Au+Au collisions for three
different energies. The $v_{1}$ values are averaged over a
centrality range of 6-55\%. In the mid-rapidity region a
significant change in slope can be seen from low to high energy
Au+Au collisions. In both 130- and 200-GeV measurements, it is
seen that, within uncertainties, $v_{1}$ is flat at mid-$\eta$, in
contrast to the 19.6-GeV results.  At all three energies, $v_{1}$
is non-zero at high $|\eta|$. Similar directed flow results have
been observed at low energy in NA49 \cite{na49} and at high energy
from STAR \cite{starv1}.

A comparison of $v_{1}$ measured at different energies in the
reference frame of one of the beam nuclei is show in
Figure~\ref{v1stack}b). The directed flow for the three energies
is similar at $\eta '\stackrel{>}{\sim}-1.5$. It should be noted
that the $v_{1}(\eta')$ measurements at the low and high energies
originate from different parts of the detector and have differing
systematic error susceptibilities and sensitivities to the
reaction plane.

For comparison with Figures~\ref{v2etaprime} and~\ref{v1stack}b),
Figure~\ref{v1stack}c) shows previously published results on the
particle multiplicity, scaled by $<Npart/2>$, as a function of
$\eta'$ for the same three energies \cite{limfrag}.

\section{Conclusions}\label{concl}

Charged hadron elliptic flow has been measured as a function of
transverse momentum and pseudorapidity for 200 GeV Au+Au
collisions. The centrality dependence of $v_{2}$ is shown over a
large range in pseudorapidity. For peripheral collisions, $v_{2}$
as a function of $|\eta|$ is not flat, even at midrapidity.

PHOBOS measurements of charged hadron directed flow for 19.6, 130
and 200 GeV Au+Au collisions are shown as a function of
pseudorapidity for $|\eta|<5.4$. Looking in the rest frame of the
nucleus, both the elliptic and directed flow show energy
independent behavior which extends to mid-rapidity.  It should be
noted that close to mid-rapidity one expects to see changes in the
curves determined as a function of pseudorapidity as opposed to
rapidity. For the elliptic flow, these changes are relatively
small (10\% at $\eta=0$) \cite{kolb}. Thus, the degree to which
the energy independence of the results extends to mid-rapidity for
the elliptic flow is intriguing.  It is difficult to reconcile
this fact with the common assumption that the particle production
at mid-rapidity differs from that in the fragmentation region,
particularly at the higher energies.

\section*{Acknowledgement(s)}
%
This work was partially supported by U.S. DOE grants
DE-AC02-98CH10886, DE-FG02-93ER40802,
DE-FC02-94ER40818,  
DE-FG02-94ER40865, DE-FG02-99ER41099, and W-31-109-ENG-38, by U.S.
NSF grants 9603486, 
0072204,            
and 0245011,        
by Polish KBN grant 2-P03B-10323, and by NSC of Taiwan Contract
NSC 89-2112-M-008-024.

\vfill\eject
\end{document}